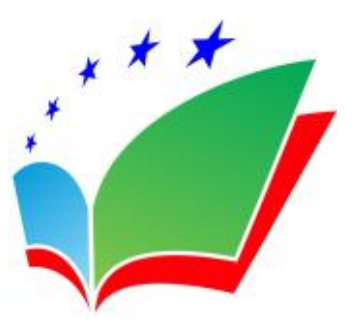

# REVITALIZING EDUCATION THROUGH ICT: A SHORT OVERVIEW OF JAPAN'S CURRENT LANDSCAPE

**Takaaki Fujita**[i]
Independent Researcher,
Japan

**Abstract:**
The domain of Information and Communication Technology (ICT) education has garnered significant consideration in recent times. However, several challenges are inherent to this area of study, including monetary expense, temporal factors, pedagogical environment, teacher training programs, incentive, syllabus design, and health-related concerns. This paper presents an analysis of the difficulties encountered in the realm of ICT education in Japan, taking into account ten different perspectives.

## 1. Introduction

### 1.1 Background of ICT education

Over the years, the integration of Information and Communication Technology (ICT) into educational pedagogy has garnered significant attention. This area encompasses pedagogical techniques that utilize digital tools such as tablet computers, Internet of Things (IoT) devices, interactive whiteboards, and personal computers. Given the importance of ICT education and its related concepts, numerous research studies have thoroughly addressed it, including its methodologies (e.g., [5,21, 27, 28, 29, 30, 31]).

The "White Paper on Education, Culture, Sports, Science and Technology (MEXT-White Paper) for FY2008" published by the Ministry of Education, Culture, Sports, Science and Technology (MEXT) emphasizes the promotion of ICT utilization, particularly the improvement of students' informational and technological literacy, efforts in programming education, and ICT training for educators. Additionally, the "Measures to Promote the Use of Advanced Technology to Support Learning in the New Era (2019)" outlines a roadmap to attain a world-class ICT-enhanced educational environment, reflecting the government's proactive approach to advancing ICT education.

The implementation of ICT education brings several benefits, including:

[i] Correspondence: email t171d603@gunma-u.ac.jp

1) Enhanced engagement and motivation among learners;
2) Improved outcomes from active learning methodologies;
3) Increased flexibility and personalization in learning;
4) Enhanced ability to assess and accommodate individual learning needs;
5) Streamlined evaluation processes, such as automatic grading;
6) Improved parent-teacher communication and collaboration;
7) Opportunities for educators to acquire ICT competencies.

Given its numerous advantages and pivotal role in the contemporary, technologically-advanced global society, the promotion and ongoing research in the domain of ICT education is crucial.

Despite the important role assigned to Information and Communication Technology (ICT) in education and the proactive measures taken by the government to encourage its adoption, the dissemination of ICT equipment in educational settings remains limited. The "White Paper on Education, Culture, Sports, Science and Technology (MEXT-White Paper)" reports that there is only one learning computer for every 5.6 students and the rate of wireless LAN penetration in classrooms is 34.5%, with high-speed internet access at 91.8%. These disparities are described as a "critical situation [24]" by the Ministry of Education, Culture, Sports, Science and Technology (MEXT).

The authors posit that the limited utilization of ICT in education is attributed to the various challenges and drawbacks unique to ICT education, such as financial and temporal costs, pedagogical environment, motivation, teacher training, and curriculum design.

### 1.2 Research Objectives of This Study

In this paper, the impediments and difficulties associated with Information and Communication Technology (ICT) education are explored and analyzed from a multitude of angles, taking into consideration not just the obstacles that may arise during the integration of ICT equipment, but also those that may present themselves during the implementation and functioning of ICT-enhanced pedagogy. In doing so, we aim to identify and address the issues inherent in current ICT education, thereby mitigating potential risks and facilitating efficient implementation of ICT technology in practical settings. For your information, the preprint of this paper has been made publicly available and can be accessed at [32].

## 2. Issues in ICT Education

The inadequate uptake of Information and Communication Technology (ICT) education is primarily attributed to the exorbitant financial and temporal expenses involved. Nevertheless, numerous other barriers and challenges impede the implementation of ICT educational initiatives, including perspectives such as:

1) Financial and temporal costs;
2) Perceptions held by educators and parents;
3) Procurement, administration, and operation of equipment and systems, including:

a. Equipment administration;
   b. Equipment procurement;
   c. Miscellaneous aspects.
4) Professional development opportunities for educators, encompassing:
   a. Technical competencies;
   b. Other considerations.
5) Curricular content, including:
   a. Course material;
   b. The aspirations of ICT education.
6) Physical learning environment;
7) Pupil motivation;
8) Health and safety;
9) Collaboration with upper management;
10) Clarity of vision and objectives.

**2.1 Maintaining the Integrity of the Specifications**

The prime obstacle hindering the implementation of ICT education is the substantial financial and temporal expenditure involved. To address this issue, particularly the financial aspect, it is imperative that a solution is devised promptly. However, it poses a significant challenge to furnish each student, especially in rural schools, with tablet terminals and personal computers, as it demands a vast budget.

Numerous sources have documented the financial and temporal costs associated with ICT education. Some of the instances are cited below.

- Sasaki (2013) highlighted that in order to establish ICT education on a national scale, the financial cost must be reduced [14].
- Kobayashi et al. (2007) enumerated the time-consuming nature of ICT device management as a concern [5].
- Takagi (2016) emphasized that teachers have limited time and resources and, as a result, cannot allocate a significant amount of time or money to ICT education [2].
- Takagi (2016) asserted that the vast budget required for ICT education and the absence of legislative adoption make it difficult to introduce suitable methods in the educational field [2][25].

It is imperative to thoroughly assess whether the benefits of ICT education justify the financial costs incurred. Despite the eagerness of school teachers to adopt ICT, the financial cost often proves to be unmanageable, and thus, sustained government support is crucial to ensure the successful implementation of an ICT environment, which requires a substantial budget. To that end, several government initiatives have been launched. Some examples are highlighted below.

- The Ministry of Education, Culture, Sports, Science, and Technology (MEXT) in its publication "Trends in the Informatization of Education (2015)" proposed support to reduce the financial cost and unit price of environmental improvements [16].

- The Ministry of Internal Affairs and Communications' Educational ICT Guidebook (2017) [17] encourages students to reduce their financial and time costs by utilizing the cloud.
- The Ministry of Education, Culture, Sports, Science, and Technology's "Measures to Promote the Use of Advanced Technology to Support Learning in the New Era (2019)" [24] presents a model for developing an ICT environment that eliminates financial costs.

**2.2 Perceptions Held by Educators and Parents**

The diffusion of ICT education can be impeded by a number of factors, including the attitudes of teachers and parents as well as the culture of the field. For instance, Toyofuku (2015) and Nemoto et al. (2016) have identified several challenges to the implementation of ICT education, including:

- A perception among some teachers and parents that learning through handwriting is more beneficial than utilizing ICT devices, due to concerns regarding the impact of ICT on children's humanity and performance, and the possibility of excessive reliance on ICT materials.
- An unrealistic expectation for the instant and profound effects of ICT education, which fails to recognize that ICT is not a panacea.
- The belief among some teachers and parents that children will be more inclined to play rather than learn when provided with ICT devices.

To address these challenges, the following approaches may be adopted:

- By presenting the numerous benefits of ICT education to teachers and parents who hold traditional beliefs, and concurrently, underscoring that not all tasks can be accomplished through ICT.
- By offering teachers hands-on experience with ICT materials, thus altering their perceptions and beliefs.

While there are numerous studies that extol the benefits of ICT education, there are also some studies that oppose its adoption, particularly as a means of instructional delivery. Urano et al. (2017) assert that the challenge lies in finding a balance between conventional and ICT-based educational methods [12].

**2.3 Procurement, Administration, and Operation of Equipment and Systems**

The implementation of ICT equipment and educational support systems often comes with its own set of challenges, including the difficulties associated with their procurement, administration, and operation. This was detailed in the "Measures to Promote the Use of Advanced Technology to Support Learning in the New Era (2019)" report released by the Ministry of Education, Culture, Sports, Science and Technology, which listed the following issues with ICT equipment:

- The provision of seamless education utilizing video conferencing and remote control is problematic.
- The maintenance, operation, and management of servers and networks entail significant financial and temporal costs.

- It is challenging for teachers to independently determine the most appropriate ICT equipment to integrate into their curriculum.
- The specifications of the proposed ICT equipment remain unclear or are inadequate.
- The acquisition of ICT equipment comes at a significantly higher cost compared to other market alternatives.

### 2.3.1 Equipment Administration

Matsunaga (2011) [5] highlights the difficulties encountered when managing a multitude of educational systems. In certain educational institutions, disparate management methods are employed to operate various systems, including personal computers for learning, tablet terminals, e-learning systems, and performance management systems. This exacerbates the management difficulties and incurs exorbitant time and financial expenses. Additionally, security and privacy policies tend to vary among each system, and the administration of IDs and passwords becomes convoluted.

To resolve these issues, the following measures could be taken:

- The creation of a manual that elucidates the procedures for attaining the goals, enabling even beginners and intermediate users to manage the system effectively.
- Instead of attempting to tackle the problem independently, the implementation of new tools and systems can help. For instance, the utilization of a password management tool that can streamline the management of passwords across all systems or the implementation of a system that can monitor the status of all systems. Although the cost may still be a consideration, free or low-cost tools are often accessible, and it would be wise to make use of them whenever possible, despite any regulatory restrictions that may exist.
- In the event that managing the systems becomes complex, the use of only the minimum number of systems is advisable [25].

### 2.3.2 Equipment Procurement

One of the challenges is the lack of equipment and infrastructure installation. This is largely owing to the financial strain, as ICT equipment tends to be cost-prohibitive. Moreover, certain apparatus has a limited-service lifespan, and may require replacement every few years [9,25].

Possible ameliorative measures include:

As an interim solution, it may be feasible to opt for equipment rental rather than out-right purchase, or to synchronize class schedules with other classes. As noted by Hara et al. (2018) [4], ICT materials are often considered supplementary. Consequently, for instance, if ICT materials are employed in only one of two or three classes, the problem can be mitigated by either procuring equipment from other schools or private organizations through rental, or by coordinating class schedules to reduce concurrent usage. Nonetheless, this is not a definitive solution.

This strategy necessitates individual families to purchase devices like laptops and tablets, which could result in an increased financial burden for families.

As highlighted in Chapter 2.1, the integration of ICT equipment requires a substantial budget, thus support from the government and local authorities is indispensable. This fiscal backing will lead to a more permanent resolution.

#### 2.3.3 Miscellaneous Aspects

Mitomo (2018) [19] posits that a reduction in communication velocity resulting from a suboptimal network infrastructure has an adverse impact on the educational outcome. As a remedy, it is recommended to minimize the number of concurrent connections.

Mori (2008) [20] contends that the e-Learning system implemented at the university has become a repository of outdated information due to the paucity of instructors capable of updating and maintaining it. To alleviate this issue, it is advisable to offer routine training opportunities.

The Ministry of Education, Culture, Sports, Science and Technology's "Measures to Promote the Use of Advanced Technology to Support Learning in the New Era (2019) [24]" highlights several points regarding the usage of data collected by Information and Communication Technology (ICT) devices. To wit, the efficacy of various equipment in utilizing the data remains unverified. Additionally, the data collected by these instruments is not applied in an actual education, and the circumstances under which the data can be used have not been taken into consideration. To address these issues, MEXT has proposed to investigate the standardization of educational data.

The limited battery life of tablets and laptops, which can only be used for two to three hours without recharging, poses another challenge that requires attention [9]. To mitigate this issue, it is suggested to take precautions to protect the devices, such as instructing students to handle them with care.

Hara et al. (2018) [4] identify copyright issues and disparities in the use of ICT across different subjects as problematic aspects of ICT teaching materials. However, with the increasing integration of ICT in subjects like physical education [3] and music [15], these disparities may be diminishing.

### 2.4. Professional Development Opportunities for Educators

#### 2.4.1 Technical Competencies

One of the difficulties associated with incorporating Information and Communication Technology (ICT) into education lies in the scarcity of teachers with sufficient technical proficiency. Nemoto et al. (2016) [7] identified that many teachers lack the ability to operate ICT equipment, making it challenging to address unexpected device behavior. The problem of the digital divide among teachers, as noted by Hara et al. (2018) [4], also adds to the complexity. The "Measures to Promote the Use of Advanced Technology to Support Learning in the New Era [24]" (2019) by the Ministry of Education, Culture, Sports, Science, and Technology also highlights the issue of teachers' unfamiliarity with selecting the appropriate ICT devices. To mitigate these difficulties, the ICT training environment for teachers must be reinforced, including through the provision of a beginner-friendly manual and user-friendly system. Additionally, providing opportunities for teachers to engage in group training and e-learning would also be beneficial. In spite of their busy schedules, it is imperative that teachers are provided with the means to expand their ICT skills.

Nagaoka et al. (2018) reported that the utilization rate of file-sharing software, a tool for sharing knowledge among schools, is relatively low [25]. To address this issue, a system should be established that enables individuals to seek the guidance of experts in cases of emergency.

#### 2.4.2 Other Considerations

Even though enhancing the technical proficiency of educators may be a necessary step, it would prove insufficient if the teachers lack intrinsic motivation. Kobayashi et al. (2007) [8] posit that teachers exhibit a disparity in both their ICT skills and motivation for ICT education. As a mitigation strategy, providing incentives such as recognition for accomplishments might prove effective. Hara et al. (2018) [4] similarly emphasize the existence of a digital divide among teachers, and suggest that holding proactive training sessions specifically geared towards novices could be a potential solution.

### 2.5 Curricular Content

#### 2.5.1 Course Material

It is imperative to revise the instructional curriculum in order to accommodate the integration of ICT education. According to Kobayashi et al. (2007) [8], a significant alteration of the academic curriculum is essential to make ICT education a crucial component of the school syllabus. This modification, however, imposes a considerable time burden on teachers.

Consequently, the availability of tools that facilitate the adaptation of class content to changes in the curriculum is crucial. For instance, a visual programming tool like Scratch proves to be beneficial, as it enables the customization of the programs taught to students, i.e., the content of the class. Microsoft Word and PowerPoint also serve as tools that can be utilized to modify the content of a lesson at any time. Nevertheless, tools that enable ease in improvement are scarce, and it would be advisable to approach companies that specialize in the development of such tools.

#### 2.5.2 The Aspirations of ICT Education

Although slightly deviating from the main topic, it is worth considering the ideal form of ICT education in correlation to the content taught in high school information classes. Tachibana et al. (2012) [6] postulate that the curriculum of information classes is overly inclined towards imparting practical skills, such as technical training in programming. In addition to technical training, there are two other essential components that ought to be emphasized in information classes:

- Theoretical comprehension of information; for instance, education in algorithms and discrete mathematics in the context of programming education. Such understanding enables students to learn how to think critically and make informed decisions when utilizing technology.

- The development of sound judgment to participate in the information society. With the widespread adoption of social network services such as Twitter, Facebook, and LINE, various problems have arisen, such as personal information leaks, slander, flaming, and excessive usage. To avoid students engaging in behaviors that could lead to these issues, it is crucial to inculcate common sense.

Unfortunately, it is a substantial challenge to find educators who possess the technical proficiency, theoretical understanding, and common sense to effectively teach all three of these elements. This is due to the scarcity of technical skills training programs for educators, including programming education. Most educators are already occupied with other responsibilities, and it would be unrealistic to expect them to acquire the skills and knowledge necessary to teach theory and common sense.

However, the author believes that the ultimate objective of information education, and by extension ICT education, should be to equip both teachers and students with the skills, theories, and common sense necessary to effectively engage with information and communication technology. All stakeholders in the educational system, including teachers, students, parents, and staff, should be encouraged to acquire these abilities, making them valuable human resources in the information age.

### 2.6 Physical Learning Environment

The utilization of Information and Communication Technology (ICT) in the educational sector is also impeded by certain environmental factors. As noted by Nemoto et al. (2016) [7], strict regulations imposed in schools, including security and privacy regulations, pose a challenge to ICT education. These regulations hinder the exploitation of data that can be garnered from ICT devices, undermining the potential benefits of ICT in education.

The Measures to Promote the Use of Advanced Technology to Support Learning in the New Era (2019) [24] established by the Ministry of Education, Culture, Sports, Science, and Technology highlights the existing imbalance between the need to secure data collected through ICT and the desire to promote its use in learning.

Additionally, Toyofuku (2015) [13] delves into the correlation between classroom teaching style and ICT education in the Japanese context. In Japan, classes are typically taught by a single teacher to all students at once, but Toyofuku (2015) [13] posits that small-group classes are more conducive to ICT education compared to large-group classes, as the latter tend to entail higher administrative costs for teachers.

### 2.7 Pupil Motivation

The interplay between ICT education and student motivation is imperative. Even if the learning environment is optimized, if students lack motivation, the educational system will go underutilized. For example, Mori (2008) [20] posits that the introduction of an e-learning system in a university failed due to the low motivation of students. Similarly, Mori et al. (2015) [10] reported negative feedback from students participating in an experiment utilizing an e-portfolio (ICT education), with complaints that the system was time-consuming, felt distant, and that they didn't feel adequately skilled. To counter these motivations, the following measures can be implemented:

- Motivate students by incorporating the recognition of their effort into their grades.
- Develop a system that seamlessly integrates the use of ICT education into the curriculum.
- Facilitate interaction and collaboration through the use of ICT systems for students with limited autonomy [22].
- Encourage teacher involvement and support for students unfamiliar with ICT devices.
- Achieve a harmonious balance between traditional and ICT-based educational methods, rather than relying solely on ICT devices [12].

### 2.8 Health and Safety

The relationship between ICT education and health problems has been a matter of concern, as outlined by the "Measures to Promote the Use of Advanced Technology to Support Learning in the New Era (2019) [24]" report issued by the Ministry of Education, Culture, Sports, Science and Technology.

One particular issue of concern is a visual impairment resulting from prolonged exposure to computer and tablet screens. As stated by Hashi et al. (2016) [18], the widespread utilization of personal computers, tablet devices, and electronic blackboards in ICT-enhanced learning has the potential to exacerbate near vision problems in children. Thus, it is imperative to implement preventative measures and accommodate the needs of children with visual impairments.

### 2.9 Collaboration with Upper Management

Kobayashi et al. (2007) [8] posit that the cooperation of top-level members is critical for fostering ICT education within an organization. As such, engaging in direct dialogue with educators is one of the methods to achieve this. Furthermore, some local authorities are making efforts to cultivate human capital that can assume a leading role in advancing ICT education.

### 2.10 Clarity of Vision and Objectives

Sasaki (2013) [14] asserts that in order to effectively propagate ICT education, it is imperative to formulate a comprehensive vision for enhancing the ICT milieu. Without a clearly defined vision, the mere integration of ICT devices may prove to be a fruitless endeavor and fall short of yielding any significant educational outcomes.

## 3. Conclusion

### 3.1 Future Tasks

Future endeavors for this scholarly work include the following:

Conducting a questionnaire survey of students and educators in the realm of education to explore and investigate new issues beyond those discussed in this paper.

Furthermore, determining the level of importance of the issues outlined in this study within the field of education.

Undertaking a comprehensive review of literature through surveying a broader range of scholarly works that pertain to the subject of ICT education, so as to foster a more nuanced and multi-faceted analysis.

Developing educational software aimed at fostering theoretical understanding and common sense regarding information, as there are currently limited educational materials that cater to these domains in comparison to those that impart technical skills.

**Acknowledgements**
I would like to express my sincere gratitude to all those who have supported me in completing this paper.

**Conflict of Interest Statement**

The author declares no conflicts of interest.

**About the Author**

Takaaki Fujita finds great enjoyment in his work as a system engineer/IT service manager, as he is constantly involved in diverse system development projects. Additionally, he delights in exploring various fields of mathematics, including discrete mathematics, combinatorics, algebra, and graph theory, alongside his keen interest in ICT education and project management. He has a master's degree in computer science.